\documentclass[runningheads]{llncs}
\usepackage{graphicx}
\usepackage{amsmath}
\DeclareRobustCommand{\bbone}{\text{\usefont{U}{bbold}{m}{n}1}}

\usepackage{algorithm} 
\usepackage{algpseudocode}
\newcommand{\algname}[1] {{\fontfamily{cmtt}\selectfont {#1}}}
\usepackage{dsfont}

\usepackage{caption}
\usepackage{subcaption}
\usepackage{multirow}
\usepackage{graphicx,wrapfig,lipsum}
\usepackage{float}
\usepackage[absolute]{textpos}
\usepackage{bbm}
\usepackage[english]{babel}
\usepackage{wrapfig}

\begin{document}
\title{Potential Factors Leading to Popularity Unfairness in Recommender Systems: A User-Centered Analysis\thanks{BNAIC/BeNeLearn Joint International Scientific Conferences on AI and Machine Learning, 2023}}
\titlerunning{Potential Factors Leading to Popularity Unfairness in Recommender Systems}

\author{Masoud Mansoury\inst{1,2}\orcidID{0000-0002-9938-0212} \and
Finn Duijvestijn\inst{1} \and
Imane Mourabet\inst{1}}
\authorrunning{Mansoury et al.}
%
\institute{University of Amsterdam, Amsterdam, Netherlands \and
Discovery Lab, Elsevier, Amsterdam, Netherlands \\
\email{m.mansoury@uva.nl}, \email{\{finn.duijvestijn,imane.mourabet\}@student.uva.nl}}
\maketitle              

\begin{abstract}
Popularity bias is a well-known issue in recommender systems where few popular items are over-represented in the input data, while majority of other less popular items are under-represented. This disparate representation often leads to bias in exposure given to the items in the recommendation results. Extensive research examined this bias from item perspective and attempted to mitigate it by enhancing the recommendation of less popular items. However, a recent research has revealed the impact of this bias on users. Users with different degree of tolerance toward popular items are not \textit{fairly} served by the recommendation system: users interested in less popular items receive more popular items in their recommendations, while users interested in popular items are recommended what they want. This is mainly due to the popularity bias that popular items are over-recommended. In this paper, we aim at investigating the factors leading to this user-side unfairness of popularity bias in recommender systems. In particular, we investigate two factors: 1) the relationship between this unfairness and users' interest toward items' categories (e.g., movie genres), 2) the relationship between this unfairness and the diversity of the popularity group in users' profile (the degree to which the user is interested in items with different degree of popularity). Experiments on a movie recommendation dataset using multiple recommendation algorithms show that these two factors are significantly correlated with the degree of popularity unfairness in the recommendation results.

\end{abstract}
\section{Introduction}

\begin{figure*}[t]
\centering
\includegraphics[width=0.75\textwidth]{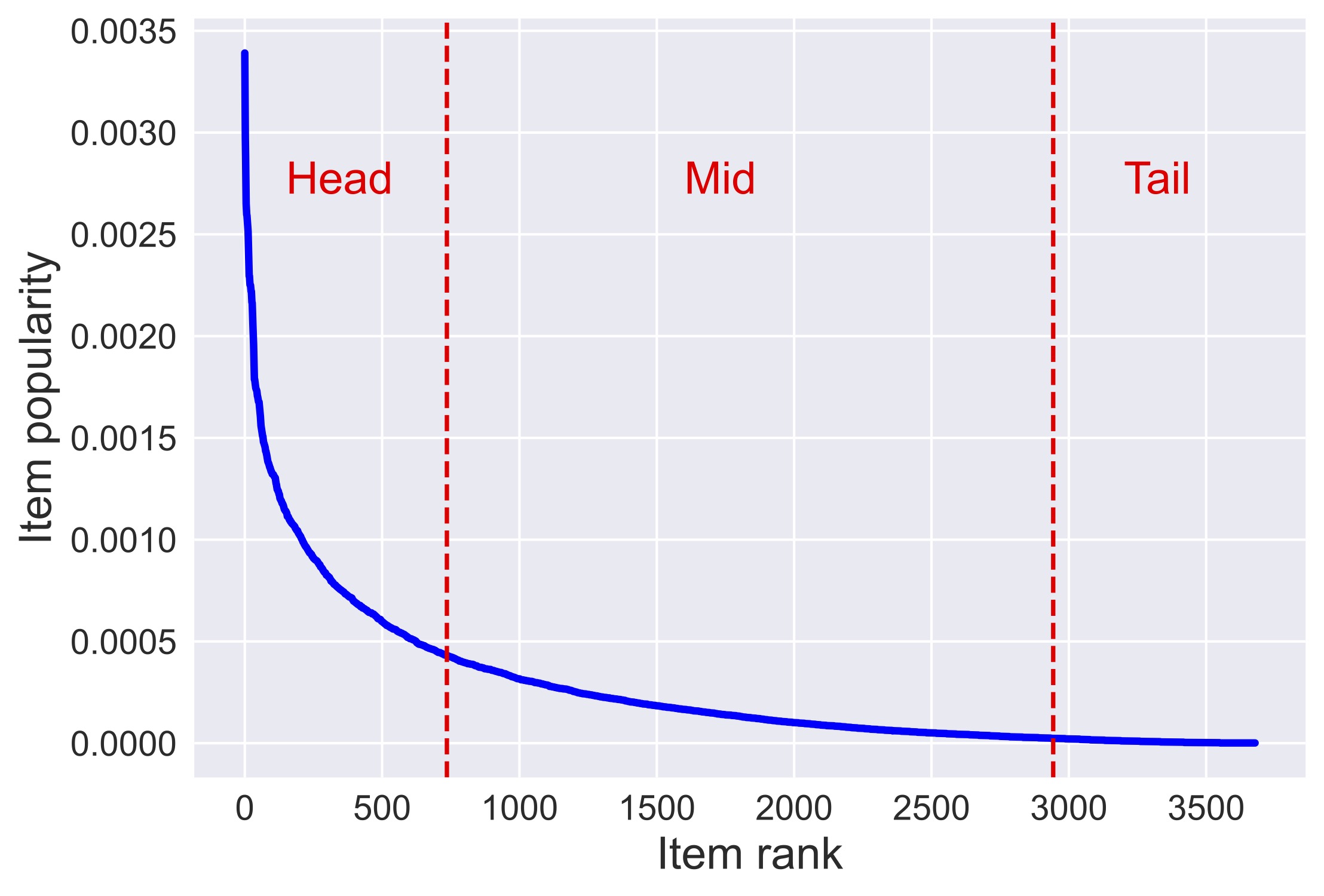}
\caption{Distribution of item popularity on MovieLens dataset. Items
are sorted based on the popularity from the most (left) to the least (right): lower item rank has higher popularity. \textit{Head} represents 20\% of most popular items, \textit{Tail} represents 20\% of least popular items, and \textit{Mid} represents the rest of 60\% items.} \label{fig_ml_pop_dist}
\end{figure*}

It is well-known that recommender systems suffer from popularity bias: a few popular items are over-represented in the rating data, while majority of other less items are under-represented \cite{abdollahpouri2017controlling,ciampaglia2018algorithmic}. Figure \ref{fig_ml_pop_dist} shows the popularity distribution of the items in MovieLens dataset. As shown, a few items with high rating frequency dominate the entire rating history. For instance, 44 most popular items (around 1\%) make up more than 10\% of the interactions. 

This bias, if not mitigated, leads to \textit{exposure bias} \cite{mansoury2022understanding,mansoury2021fairness} where popular items are over-recommended, while unpopular items are under-recommended in the recommendation results. This unfairness in the representation of items in the recommendation lists also leads to unfairness against suppliers who provided those items in the system \cite{abdollahpouri2020unfair,surer2018multistakeholder,mansoury2021graph}. Also, due to the \textit{feedback loop} phenomena in which users profiles get updated over time via potentially biased recommendations, popularity bias can be amplified over time \cite{mansoury2020feedback,sinha2016deconvolving,mansoury2023fairness}. Mansoury et al. in  \cite{mansoury2020feedback} showed that in long run, as users interact with the system, amplification of popularity bias not only intensifies the unfairness, but also degrades the performance of the recommendations.

Most existing works study the popularity bias from the aforementioned item perspective \cite{rastegarpanah2019fighting,zhu2021popularitya,steck2019collaborative}. Various algorithms are developed to mitigate this bias by promoting less popular items, known as \textit{long-tail}, in the recommendation results to compensate for their under-recommendation and increase their visibility \cite{beutel2019fairness,yao2017beyond,zhu2021popularity}. For example, Zhu et al. in \cite{zhu2021popularity} proposed the idea of popularity compensation as a post-processing approach that adjusts the predicted scores for items according to their popularity: increasing the score for less popular items to promote them in the recommendation list; Abdollahpouri et al. in \cite{abdollahpouri2017controlling} proposed an in-processing approach that mitigates the popularity bias by incorporating a regularization term into the objective function. While these algorithms effectively mitigate item-side popularity bias, they ignore the user-centered impact of popularity bias.

Recent research has shown that recommendation models fail to properly follow the users' interest toward popular items \cite{abdollahpouri2019unfairness,abdollahpouri2021user}. Abdollahpouri et al. in \cite{abdollahpouri2019unfairness} demonstrated that popularity bias can lead to unfairness among users. They showed that users differ in their interest toward popular items: some users have high interest to popular items (blockbuster-focused), while some users have high interest to less popular items (niche-focused). 
Due to the popularity bias in input data and over-recommendation of these items in the recommendation results, these user groups are unfairly treated. This means that the impact of popularity bias is not equal across users belonging to different groups. Niche-focused users, while are interested in less popular items, receive popular items in their recommendation list, leading to a larger impact on these users.

The same patterns were found later by Kowald et al. \cite{kowald2020unfairness} on the music domain where users with a lesser interest in mainstream songs were impacted more severely by the
popularity bias. These findings showed that popularity bias impacts different users differently and hence it is important to take users into account when mitigating this bias. To address this issue, abdollahpouri et al. in \cite{abdollahpouri2021user} proposed the idea of \textit{popularity calibration}, a post-processing approach, that generates the recommendations to be approximately matched or calibrated with the interest of users toward popularity spectrum in their profile.

In this paper, we base our analysis on findings in \cite{abdollahpouri2019unfairness,abdollahpouri2021user} and aim at investigating the potential factors that lead to user-centered unfairness of popularity bias. We in particular investigate two factors: 1) the impact of \textit{genre popularity} on this popularity unfairness, 2) the impact of \textit{popularity diversity}, the diversity of users' profile in terms of their interest to various item groups (e.g., popular and unpopular), on the degree of this user-centered popularity unfairness. Experiments on a real-world dataset show that our identified factors are highly correlated with the degree of unfairness in recommendation results, suggesting ideas for future research to alleviate this unfairness against users.

\section{The unfairness of popularity bias}\label{sec_pop_unfairness}

Throughout the paper, we use MovieLens 1M dataset \cite{harper2015movielens} for our analysis. In this dataset, 6,040 users provided 1,209,000 ratings on 3,706 items (movies). Each item is associated with one or more genres and there are in total 18 different genres in the dataset. Figure \ref{fig_ml_pop_dist} shows the popularity distribution of MovieLens dataset. Vertical axis shows the popularity of the items and horizontal axis shows the items sorted by their popularity value in descending order. It is evident that the distribution is long-tailed and the data heavily suffers popularity bias.

In this long-tail distribution, we define popular items, also referred to as \textit{Head} items (\textit{H}), as the top 20\% of most popular items (items frequently interacted in the rating data). The rest of the less popular items are often referred to as \textit{long-tail} items and can be further divided into: \textit{Tail} items (\textit{T}) as the top 20\% of the least popular items (items rarely interacted in the rating data), and \textit{Mid} items (\textit{M}) are the ones in between that includes 60\% of the rest of the items. 

Following the findings in \cite{abdollahpouri2019unfairness}, popularity bias in rating data leads to discrimination against users: users interested to popular items are much better served by the recommender systems than the ones interested in less popular items. In other words, users interested in less popular items, while expected to get unpopular items in their recommendation lists, receive more popular items as the recommendation. In the following section, we briefly review this issue.

\subsection{User grouping in terms of propensity to item popularity}\label{sec_user_group_item}

We group users based on their interest toward popular items. Users can be mostly interested in either popular items, unpopular items, or items in the middle of popularity spectrum. Hence, we define three groups of users:

\begin{itemize}
    \item \textbf{Blockbuster-focused users:} Users with interest to popular items.
    \item \textbf{Diverse taste users:} Users with diverse interest toward both popular and less popular items. 
    \item \textbf{Niche-focused users:} Users with interest to less popular items.
\end{itemize}

Given users' profile with items they interacted with, we first derive the ratio of \textit{H}, \textit{M}, and \textit{T} items in their profile. For example, if a user interacted with 7 items in \textit{H}, 2 items in \textit{M}, and 1 item in \textit{T}, then the ratios for \textit{H}, \textit{M}, and \textit{T} in her profile would be 0.7, 0.2, 0.1, respectively. Then, we follow this process for grouping the users: 1) extract 20\% of the users with the highest ratio of \textit{T} items in their profile to form \textbf{Niche group} (i.e., sort the users based on the ratio of \textit{T} items in descending order and return top 20\%), 2) extract 20\% of the users with the highest ratio of \textit{H} in their profile to form \textbf{Blockbuster group} (i.e., sort the users based on the ratio of \textit{H} items in descending order and return top 20\%), 3) and finally, extract users who are neither in niche-focused nor in blockbuster-focused groups to form \textbf{Diverse group}.

\subsection{Evaluation metrics}

For measuring the degree of popularity unfairness in the recommendation results, we use metrics introduced in \cite{abdollahpouri2020multi,abdollahpouri2019unfairness,abdollahpouri2021user}.

\noindent\textbf{Popularity Lift (PL):} This is a measure of popularity amplification of recommendation list delivered to the user with respect to the popularity of her profile. We denote $APP_u$ as the \textit{Average Profile Popularity} of user $u$'s profile and $ARP_u$ as the \textit{Average Recommendation Popularity} of $u$'s recommendations, both computed as

\begin{equation*}
    APP_u=\frac{\sum_{i \in \mathcal{R}_u}{pop(i)}}{|\mathcal{R}_u|} \;, \;\;\;\;\;\;\;\;\;
    ARP_u=\frac{\sum_{i \in \mathcal{L}_u}{pop(i)}}{|\mathcal{L}_u|}
\end{equation*}

\noindent where $pop(i)$ returns the popularity of item $i$ as fraction of interactions on $i$, $\mathcal{R}_u$ is the items that $u$ interacted, and $\mathcal{L}_u$ is the items recommended to $u$. Thus, \textit{PL} of user $u$ is computed as

\begin{equation}
    PL_u = \frac{ARP_u-APP_u}{APP_u}
\end{equation}

Positive value for \textit{PL} signifies that the recommendation list of a user is on average more popular than his profile, while negative value indicates that the user's profile is on average more popular than her recommendation list. However, $PL_u=0$ means no popularity amplification or degradation which is more desired. We further define the popularity lift for a group $G \in \{\text{Blockbuster}, \text{Diverse}, \text{Niche}\}$ as $PL_G=\frac{ARP_G-APP_G}{APP_G}$ where $APP_G=\frac{\sum_{u \in G}{APP_u}}{|G|}$ and $ARP_G=\frac{\sum_{u \in G}{ARP_u}}{|G|}$.

\noindent\textbf{User Popularity Deviation (UPD):} This is a measure of \textit{popularity miscalibration} \cite{abdollahpouri2021user} in the recommendation results. In other words, given $P_u$ and $Q_u$ as the ratio of each item group in $u$'s profile and recommendation list, respectively, $UPD$ calculates the distance between $P_u$ and $Q_u$, indicative of how well $Q_u$ is calibrated with respect to $P_u$. For example, if a user's profile represents the ratios of $\langle 0.7,0.2,0.1\rangle$ for $\langle H,M,T\rangle$, a recommendation list is considered to be perfectly calibrated (i.e., $UPD=0$) if it also consists of the same ratios. 
Given $C=\{H,M,T\}$, we calculate the propensity of each user $u$ towards
each item group $c \in C$ in her profile as:

\begin{equation}
    p_u(c) = \frac{\sum_{i \in \mathcal{R}_u}{\bbone{}(i \in c)}}{\sum_{c' \in C}{\sum_{i \in \mathcal{R}_u}{\bbone{}(i \in c')}}}
\end{equation}

\noindent and the ratio of each item group $c \in C$ in her recommendation as:

\begin{equation}
    q_u(c) = \frac{\sum_{i \in \mathcal{L}_u}{\bbone{}(i \in c)}}{\sum_{c' \in C}{\sum_{i \in \mathcal{L}_u}{\bbone{}i \in c'}}}
\end{equation}

Thus, we define $P_u=\{p_u(H),p_u(M),p_u(T)\}$ and $Q_u=\{q_u(H),\allowbreak q_u(M),q_u(T)\}$. Now, we calculate $UPD_u$ as:

\begin{equation}
    UPD_u=\mathcal{J}(P_u,Q_u)
\end{equation}

\noindent where $\mathcal{J}(p_u,q_u)$ is Jensen–Shannon
divergence \cite{lin1991divergence} between $p_u$ and $q_u$ which measures the statistical distance between two probability. Similar to $PL$, we calculate $UPD_G$ for a group $G$ (e.g., $\textbf{Niche}$) as the average $UPD$ of all users belonging to $G$.

\subsection{Experimental setup}

For experiments, we split the users' profile in MovieLens dataset into 80\% as training set and 20\% as test set. The training set is used to build the recommendation model and generating the recommendations for each user. The test set is used for evaluating the performance of the recommendations. We perform the experiments using the following recommendation algorithms: Bayesian Personalized Ranking (\algname{BPR})~\cite{rendle2009bpr}, 
Biased Matrix Factorization (\algname{BiasedMF})~\cite{koren2009matrix}, 
User-based Neighborhood Model (\algname{UserKNN})~\cite{resnick1994grouplens}, and Item-based Neighborhood Model (\algname{ItemKNN})~\cite{sarwar2001item}. We also use two non-personalized algorithms: most popular item recommendation (\algname{Popular}) and random item recommendation (\algname{Random}). We tune each algorithm by performing a grid-search over the hyperparameter space and reporting the best-performing results in terms of precision, recall, and ranking quality (nDCG) of the recommendations. 
In recommendation generation, a recommendation list of size 10 is generated for each user. We use \textit{Librec-Auto}~\cite{mansoury2019algorithm} for running our experiments.

\begin{table}[t]
\centering
\setlength{\tabcolsep}{7pt}
\captionof{table}{Performance of recommendation algorithms on MovieLens dataset in terms of accuracy metrics.} \label{tab_res}
\begin{tabular}{lrrr}
\hline
\hline
 algorithms & $precision@10$ & $recall@10$ & $nDCG@10$ \\
 \hline
 \algname{BPR} & 0.183 & 0.072 & 0.184 \\
 \algname{BiasedMF} & 0.108 & 0.039 & 0.111 \\
 \algname{ItemKNN} & 0.315 & 0.130 & 0.310 \\
 \algname{UserKNN} & 0.243 & 0.095 & 0.243 \\
 \algname{Popular} & 0.186 & 0.068 & 0.185 \\
 \algname{Random} & 0.009 & 0.003 & 0.007 \\
\hline
\hline
\end{tabular}
\vspace{-10pt}
\end{table}

\begin{figure*}[t]
    \centering
    \begin{subfigure}[b]{0.85\textwidth}
        \includegraphics[width=\textwidth]{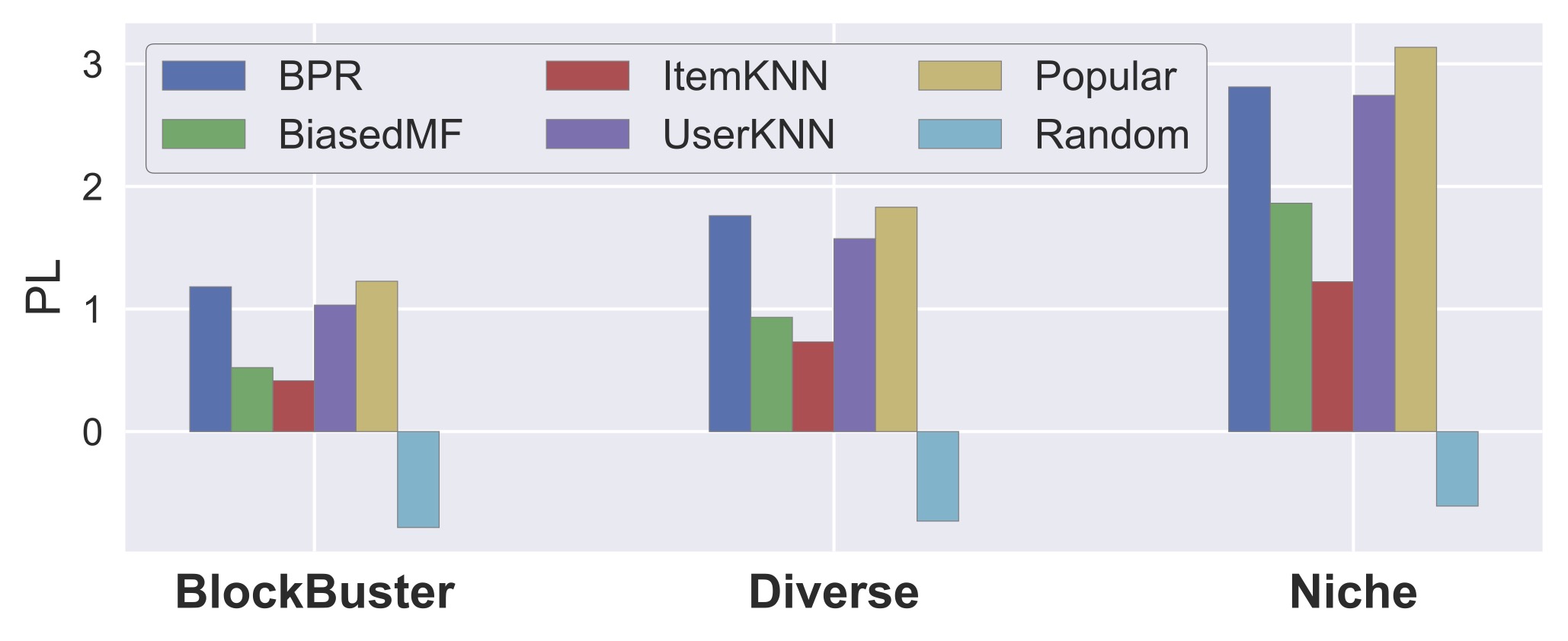}
    \end{subfigure}
\caption{$PL$ of different recommendation algorithms across various user groups on MovieLens dataset.}\label{fig_ml_pl_group}
\end{figure*}
\begin{figure}[t!]
    \centering
    \begin{subfigure}[b]{0.85\textwidth}
        \includegraphics[width=\textwidth]{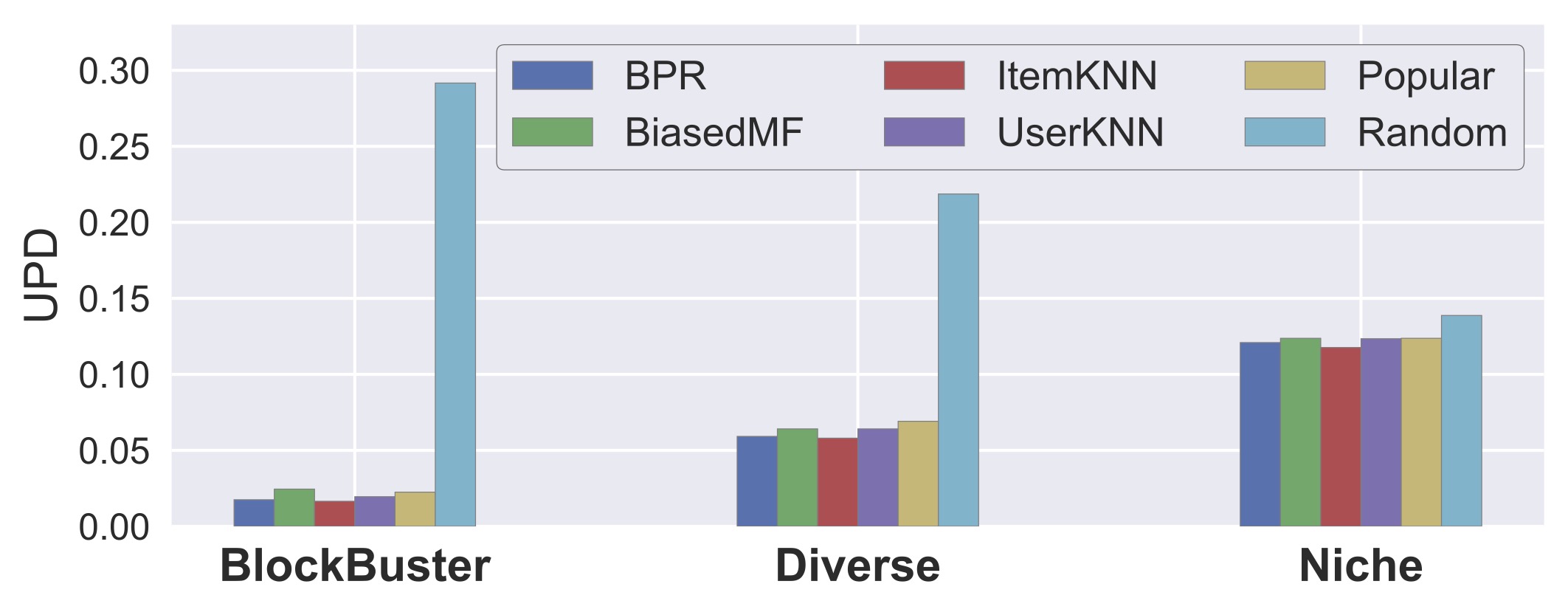}
    \end{subfigure}
\caption{$UPD$ of different recommendation algorithms across various user groups on MovieLens dataset.}\label{fig_ml_upd_group}
\end{figure}

\subsection{Experimental results}\label{sec_exp_item_pop}

Table \ref{tab_res} presents the performance of the recommendation models based on accuracy metrics. These results suggest that the models are accurate (reliable) enough to be used for the rest of analysis in this paper.

Figure \ref{fig_ml_pl_group} shows the \textit{PL} of the recommendation models for each group of users. Note that the goal here is not to compare the recommendation models, but instead to compare the performance of each recommendation model across different groups of users. As shown, all recommendation models (except \algname{Random}) consistently reveal this pattern: $PL^{Niche}\succ PL^{Diverse}\succ PL^{Blockbuster}$. This indicates that users in Niche-focused group are getting worst recommendation in terms of their interest toward popular items, and users in Blockbuster-focused group are served the best (lowest $PL$).

Figure \ref{fig_ml_upd_group} shows the performance of recommendation models in terms of $UPD$ across different groups of users. Similarly, $UPD$ of Niche-focused group is the highest, while it is the lowest for Blockbuster-focused group (except for \algname{Popular}). This shows that the recommendation models do not consistently follow the popularity interest of different users, indicative of the user-side unfairness.  

\section{Profile Inconsistency: propensity for genre popularity vs. item popularity}

\begin{figure*}[t]
    \centering
    \begin{subfigure}[b]{\textwidth}
        \includegraphics[width=\textwidth]{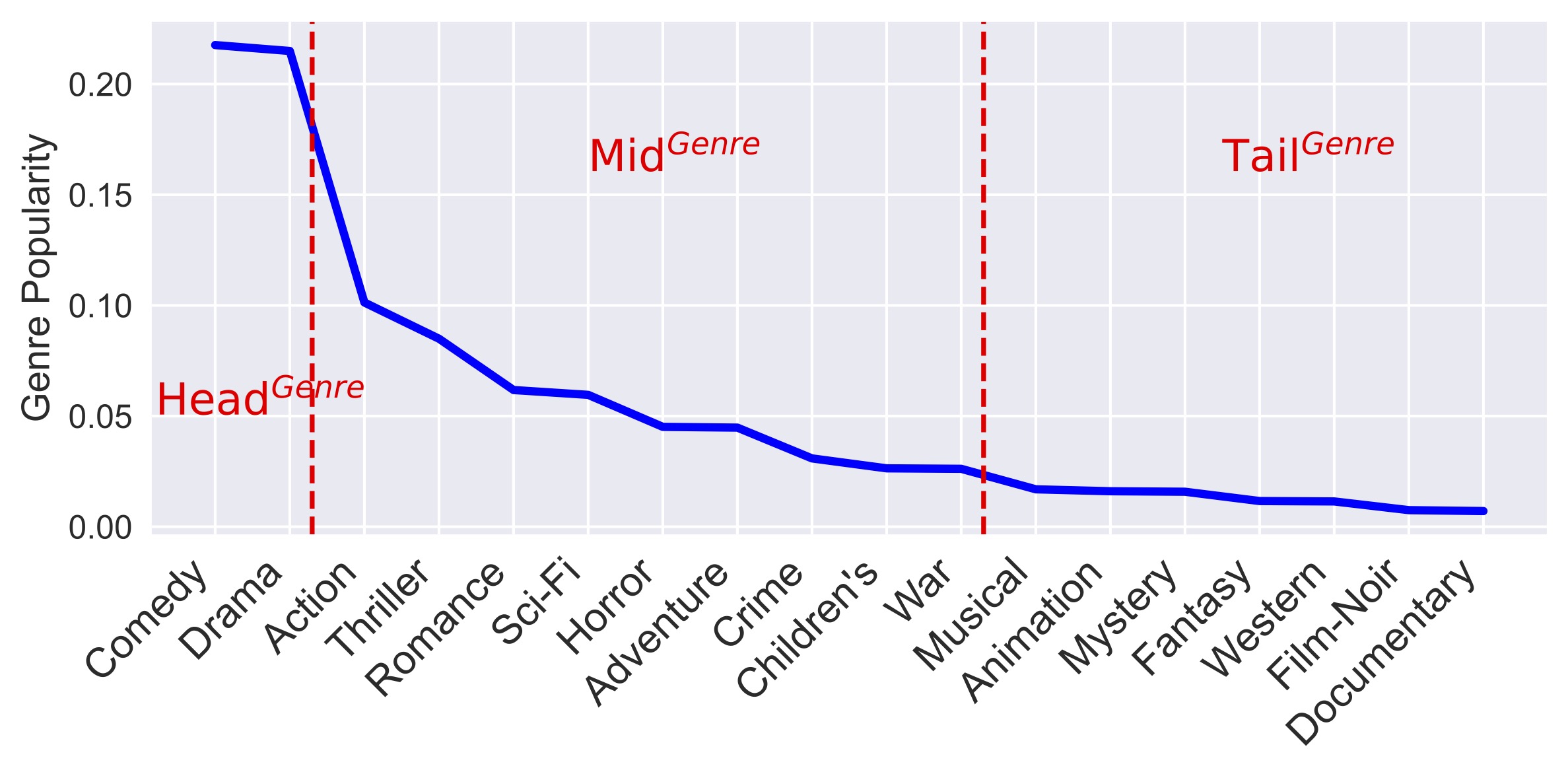}
    \end{subfigure}
\caption{Distribution of genre popularity on MovieLens dataset.}\label{fig_ml_pop_dist_genre}
\end{figure*}

\subsection{Genre popularity}\label{sec_genre_pop}

Figure \ref{fig_ml_pop_dist_genre} shows the distribution of genre popularity in MovieLens dataset. Since movies are assigned multiple genres, we equally distribute the interaction on a movie to its genres. For example, if a movie is assigned Action and Drama genres, for computing the genre popularity, we count the interaction on this movie for its genres as 0.5 for Action and 0.5 for Drama. Analogous to the distribution of item popularity, the distribution of genre popularity is long-tailed, meaning that a few genres (Comedy and Drama) make up more than 40\% of the interactions in the rating data. 

We perform the same process described in section \ref{sec_pop_unfairness} for grouping the genres and the users in terms of genre popularity. For grouping the genres, we define the most popular genres as $Head^{Genre}$, the least popular genres as $Tail^{Genre}$, and the rest of the genres as $Mid^{Genre}$. These genre groups are shown in Figure \ref{fig_ml_pop_dist_genre}. For grouping the users, we follow the process described in section \ref{sec_user_group_item}. We first extract 20\% of users with the highest interest to genres in $Tail^{Genre}$ to form $\textbf{Niche}^{Genre}$. Then, among the rest of the users, we extract 20\% of them with highest interest to $Head^{Genre}$ to create $\textbf{Blockbuster}^{Genre}$. Finally, we form group $\textbf{Diverse}^{Genre}$ with the rest of the users. To distinguish item and genre grouping in this section, we refer to item grouping as $Head^{Item}$, $Mid^{Item}$, and $Tail^{Item}$. We also refer to item-based user groups (section~\ref{sec_user_group_item}) as $\textbf{Blockbuster}^{Item}$, $\textbf{Diverse}^{Item}$, and $\textbf{Niche}^{Item}$.

\begin{figure*}[t]
    \centering
    \begin{subfigure}[b]{0.8\textwidth}
        \includegraphics[width=\textwidth]{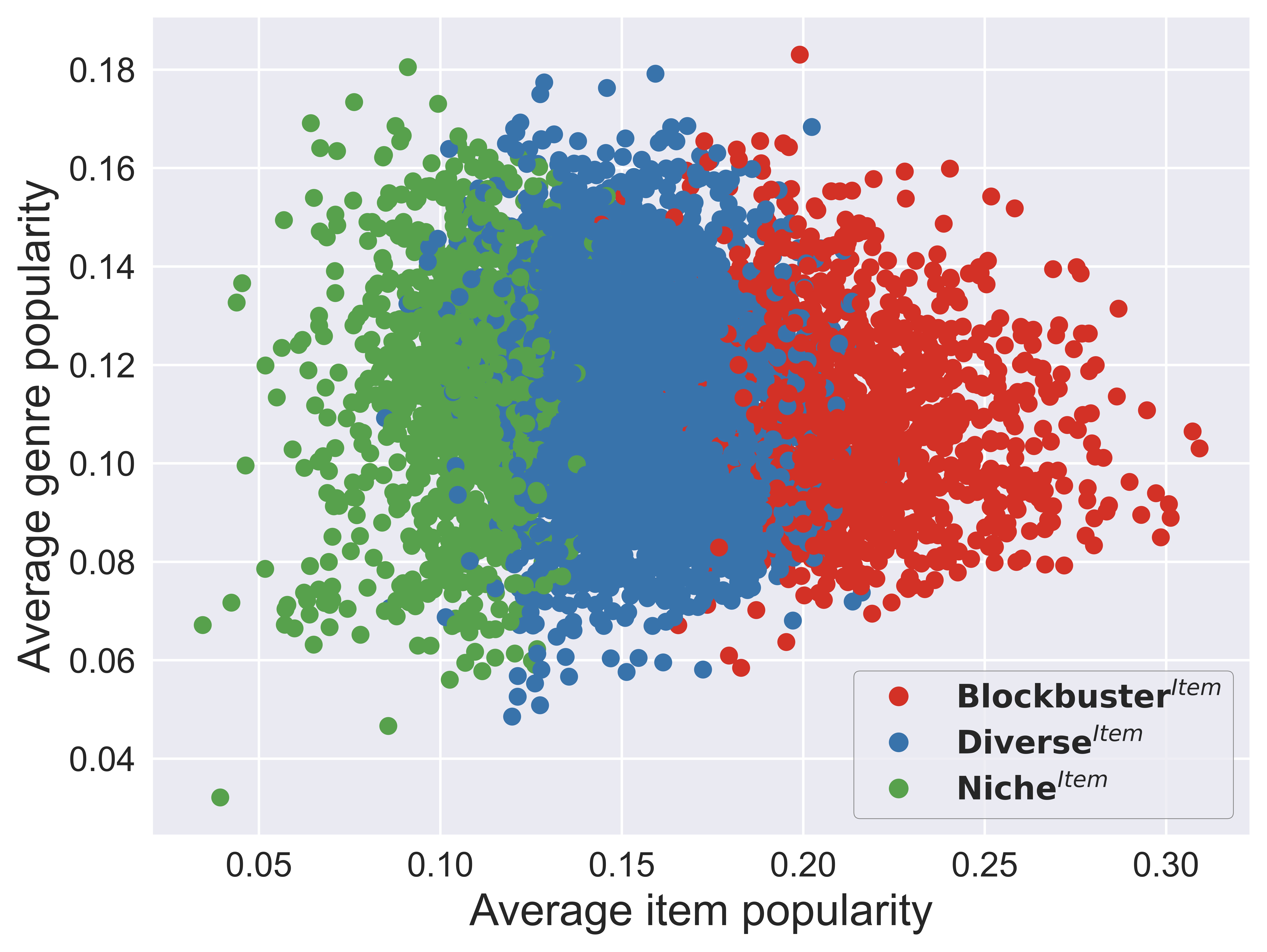}
    \end{subfigure}
\caption{The relationship between the average item popularity and the average genre popularity of users' profile. Colors show user groups based on item popularity (section~\ref{sec_user_group_item}).}\label{fig_ml_pop_item_genre}
\end{figure*}

\subsection{The relationship between the propensity to item popularity and genre popularity}

Figure \ref{fig_ml_pop_item_genre} shows the relationship between the average item popularity and average genre popularity of each user's profile. Each point in the plot represents a user, x-axis represents its average item popularity, and y-axis represents its average genre popularity. As shown, there is no correlation between users' interest toward item popularity and genre popularity. This indicates that in terms of item popularity, a user might be interested in popular items, while in terms of genre popularity, she might be interested in unpopular genres. Therefore, this finding gives us clue for further investigation on utilizing genre information to alleviate the unfairness against Niche-focused users.

\subsection{The impact of genre popularity on disparate treatment}\label{sec_impact_genre_pop}

\begin{figure*}[t]
    \centering
    \begin{subfigure}[b]{0.9\textwidth}
        \includegraphics[width=\textwidth]{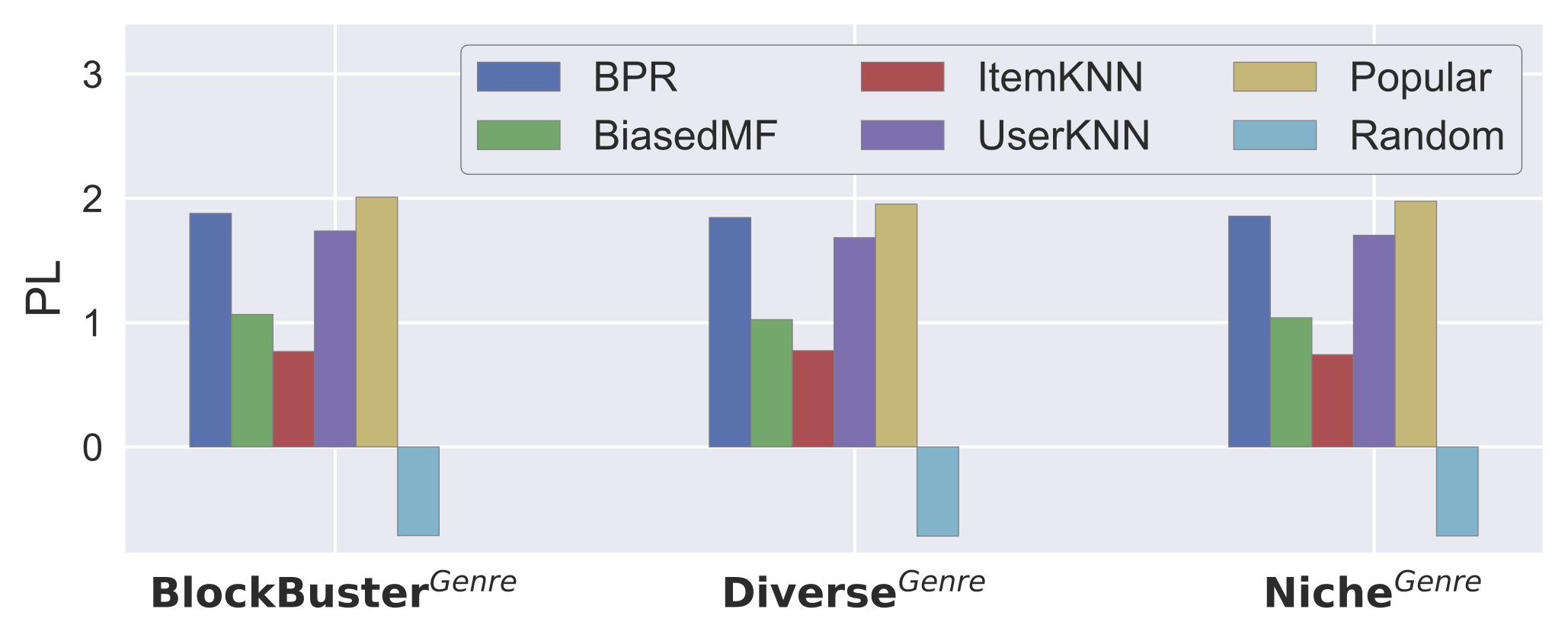}
    \end{subfigure}
\caption{$PL$ of different recommendation algorithms across various genre-based user groups on MovieLens dataset.}\label{fig_ml_pl_genre_group}
\end{figure*}

Given $\textbf{Blockbuster}^{Genre}$, $\textbf{Diverse}^{Genre}$, and $\textbf{Niche}^{Genre}$ as user groups based on users interest toward genre popularity (described in section \ref{sec_genre_pop}), the goal in this section is to reproduce the results in section \ref{sec_exp_item_pop} for these user groups. Figure \ref{fig_ml_pl_genre_group} shows the $PL$ of recommendation models for the genre-based users' groups. Again, the aim is not to compare the performance of recommendation models, but instead to compare the behavior of a recommendation model across different groups of users. 

\begin{figure*}[t]
    \centering
    \begin{subfigure}[b]{0.9\textwidth}
        \includegraphics[width=\textwidth]{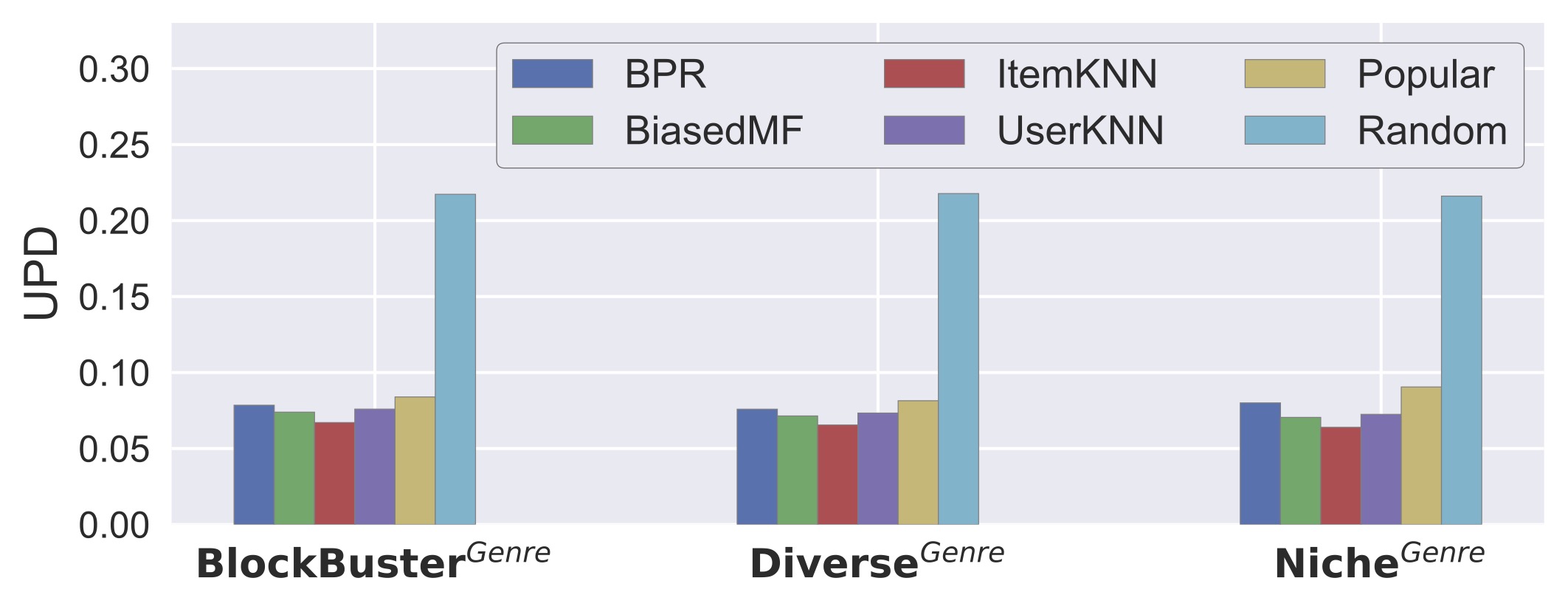}
    \end{subfigure}
\caption{$UPD$ of different recommendation algorithms across various genre-based user groups on MovieLens dataset.}\label{fig_ml_upd_genre_group}
\end{figure*}

Unlike the results obtained for item-based user groups in Figure \ref{fig_ml_pl_group}, Figure \ref{fig_ml_pl_genre_group} shows no unfairness among different groups of users. This means that niche-focused users are equally served as blockbuster-focused users by different recommendation models. The same pattern can also be observed in terms of $UPD$ in Figure \ref{fig_ml_upd_genre_group}.

\subsection{Profile inconsistency}

Given the finding in section \ref{sec_impact_genre_pop}, we are interested in investigating how much the user groups created based on item popularity (section \ref{sec_user_group_item}) match those based on genre popularity (section \ref{sec_genre_pop}). We define the percentage of overlap between group $A$ and $B$ as $A \wedge B=\frac{|A \cap B|}{|A|}\times 100$ where $A \cap B$ is the intersection of $A$ and $B$. 

\begin{figure*}[t]
    \centering
    \begin{subfigure}[b]{0.95\textwidth}
        \includegraphics[width=\textwidth]{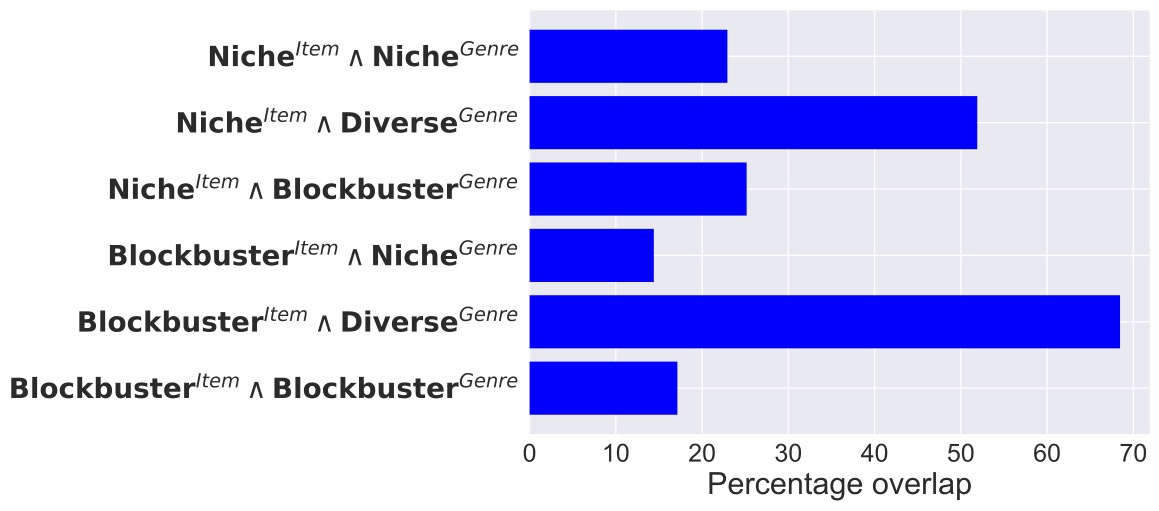}
    \end{subfigure}
    \vspace{-10pt}
\caption{Percentage of overlapped users in item-based and genre-based user groups.}\label{fig_ml_inconsistency_perc}
\end{figure*}

Figure \ref{fig_ml_inconsistency_perc} shows the percentage of overlap between item-based and genre-based user groups. 
First, it shows that the main overlap happens with $\textbf{Diverse}^{Genre}$ group which can be expected as it is the group close to both $\textbf{Niche}$ and $\textbf{Blockbuster}$. Comparing $\textbf{Niche}^{Item}$ with genre-based users' groups, only around 23\% of users in $\textbf{Niche}^{Item}$ are also in $\textbf{Niche}^{Genre}$, but surprisingly around 25\% of users in $\textbf{Niche}^{Item}$ are in $\textbf{Blockbuster}^{Genre}$. Similarly, comparing $\textbf{Blockbuster}^{Item}$ with genre-based user groups, it shows that around 17\% and 14\% of users in $\textbf{Blockbuster}^{Item}$ are in $\textbf{Blockbuster}^{Genre}$ and $\textbf{Niche}^{Genre}$, respectively.

With observations from Figure \ref{fig_ml_inconsistency_perc}, we define \textit{profile inconsistency} as the degree to which the item popularity of a user's profile does not match its genre popularity. Thus, we first define a binary function $inconsistent(i)$ that returns 1 if $i$'s group is inconsistent with its genres' groups. For example, for $i$ in $Tail^{Item}$, we say $i$ is consistent ($inconsistent(i)=0$) if it only belongs to $Tail^{Genre}$, otherwise we say $i$ is inconsistent ($inconsistent(i)=1$). Now, we define the profile inconsistency ($PI$) of each user $u$ as follows:

\begin{equation}
    PI_u=\frac{\sum_{i \in \mathcal{R}_u}{inconsistent(i)}}{|\mathcal{R}_u|}
\end{equation}

\noindent $PI_u$ computes the ratio of the \textit{inconsistent} items that user interacted with. Lower $PI_u$ implies that a user's interest toward popularity spectrum (either popular or less popular) is stronger, meaning that the items $u$ interacted with and their genres are in the same popularity group. On the other hand, higher $PI_u$ indicates that assigned group to $u$ is not sufficiently reliable and $u$ might also have certain degree of interest toward other groups.

\begin{figure*}[t]
    \centering
    \begin{subfigure}[b]{0.74\textwidth}
        \includegraphics[width=\textwidth]{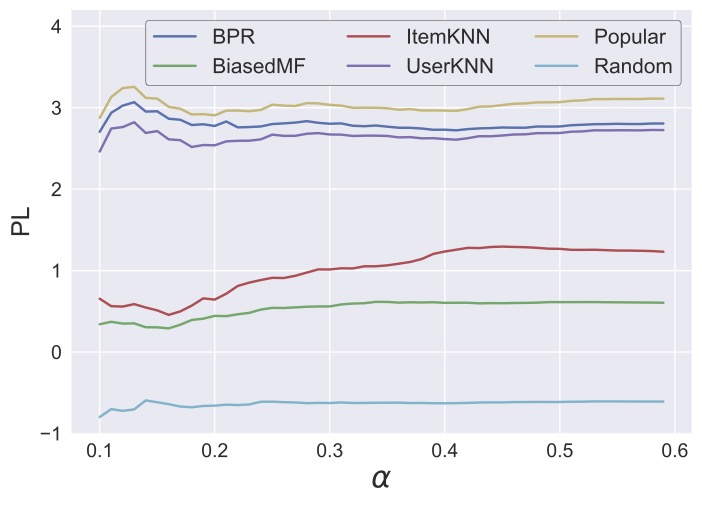}
    \end{subfigure}
    \vspace{-10pt}
\caption{The relationship between profile inconsistency and $PL$ of different recommendation models for $\textbf{Niche}^{item}$ users on MovieLens dataset.}\label{fig_ml_inconsistency_pl}
\end{figure*}
\begin{figure}[t!]
    \vspace{-10pt}
    \centering
    \begin{subfigure}[b]{0.74\textwidth}
        \includegraphics[width=\textwidth]{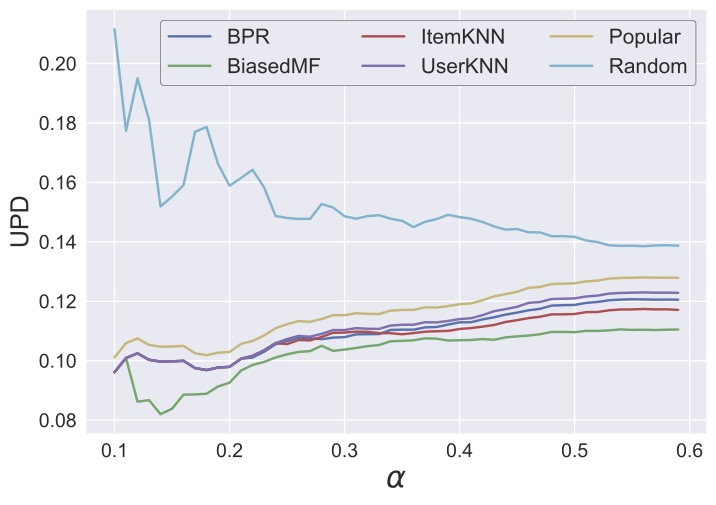}
    \end{subfigure}
    \vspace{-10pt}
\caption{The relationship between profile inconsistency and $UPD$ of different recommendation models for $\textbf{Niche}^{item}$ users on MovieLens dataset.}\label{fig_ml_inconsistency_upd}
\end{figure}

Now, the question is: \textit{do the recommendation models behave fairer for consistent users?} To answer this question, among users in $\textbf{Niche}^{Item}$, we look at the degree of unfairness for those users who their $PI$ is lower than a threshold. Given a threshold $\alpha$, we extract users who their $PI$ is less than $\alpha$ and then we report their average $PL$ and $UPD$. The reason we only look at $\textbf{Niche}^{Item}$ is that this group is the \textit{protected group} in our analysis and the group experiencing the highest disparity in the recommendation results. 

Figure \ref{fig_ml_inconsistency_pl} shows the $PL$ of the recommendation models for a subset of users in $\textbf{Niche}^{Item}$ with $PI_u \leq \alpha$ for $\alpha \in [0.1,0.6]$. The results show that $PL$ of factorization models (\algname{BPR} and \algname{BiasedMF}) slightly improve for more consistent users and as $\alpha$ increases (more inconsistent users are included), $PL$ also increases (higher disparity). This pattern cannot be observed for neighborhood models (\algname{UserKNN} and \algname{ItemKNN}). However, this pattern is stronger for $UPD$ across all algorithms as shown in Figure \ref{fig_ml_inconsistency_upd}. In all recommendation models (except \algname{Random}), with increasing $\alpha$ (more inconsistent users), $UPD$ also increases.
These results suggest ideas for improving the fairness of the recommendations using genre information: incorporating the genre information when generating recommendations, particularly for users with high profile inconsistency. 

\begin{figure*}[t]
    \centering
    \begin{subfigure}[b]{\textwidth}
        \includegraphics[width=\textwidth]{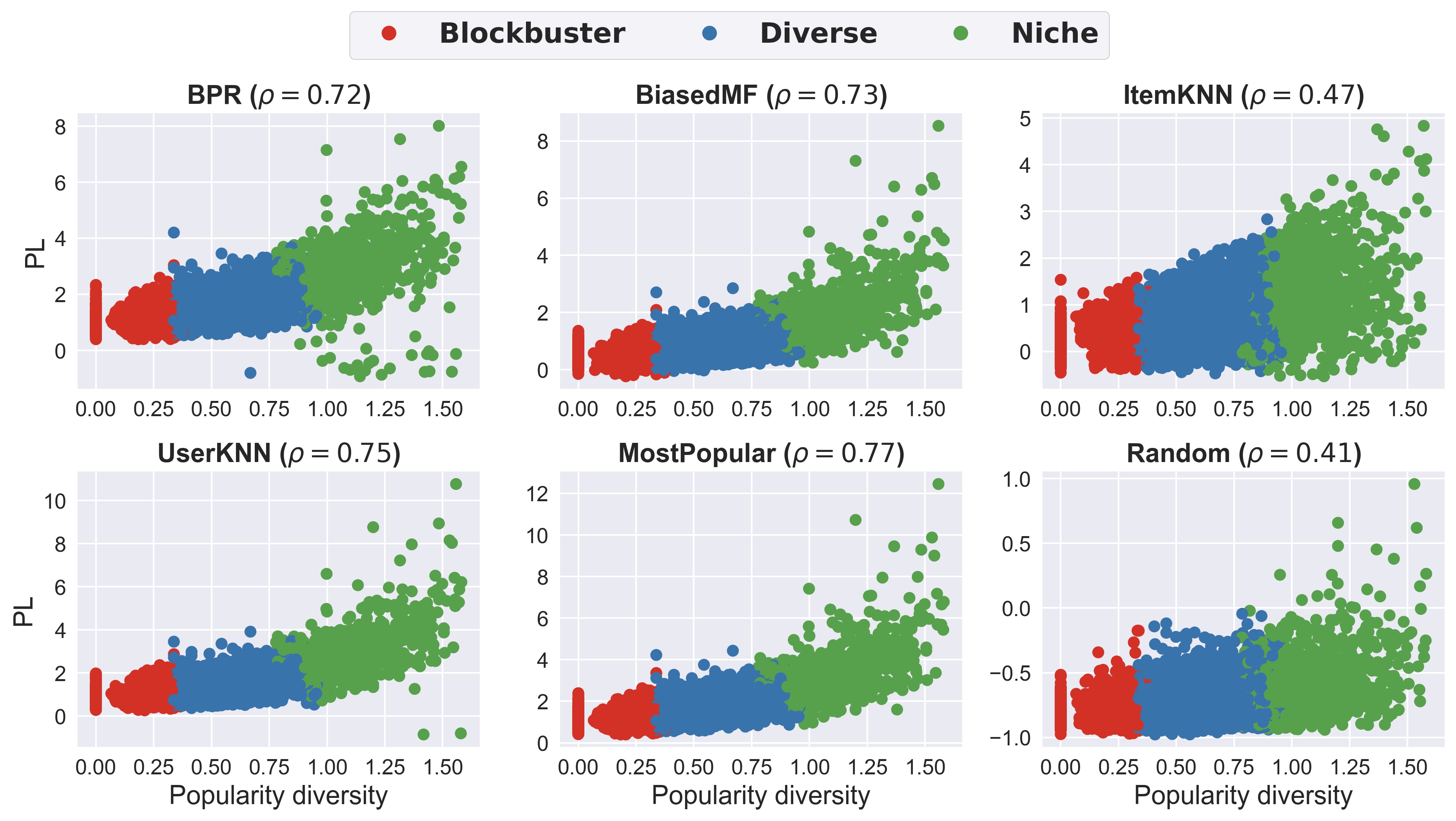}
    \end{subfigure}
\caption{The relationship between the popularity diversity of users' profile and $PL$ of their recommendations on MovieLens dataset. $\rho$ is the correlation value of these two variables.}\label{fig_ml_pop_entropy_pl}
\end{figure*}

\section{Popularity Diversity of users profile}

The second factor that we investigate is the \textit{popularity diversity} of the users' profile and we aim at showing how correlated this factor is with the ability of the recommendation model to follow the users' interest toward popularity spectrum. Popularity diversity refers to the fact that how interacted items in a user's profile are from diverse item groups. For example, a profile with items from $\langle Head, Head, Mid, Tail \rangle$ is considered to be more diverse than a profile with items from $\langle Head, Head, Head, Head \rangle$ as it covers more diverse sets of item groups. Therefore, if a user's profile has lower popularity diversity, it can be interpreted as more focused the user is in interacting with the items. For example, when a user only interacted with $Head$ items, it means that he/she has high concentration on popular items and his/her profile would result in low popularity diversity. 

For computing the popularity diversity of a user's profile, we first extract the item groups corresponding to the interacted items and then, we compute \textit{entropy} on the vector of extracted item groups. This way, a profile where every interacted item belongs to one item group would result in the lowest entropy and consequently the lowest popularity diversity. And vice versa, a profile where interacted items belong to various item groups would represent higher popularity diversity. 

\begin{figure*}[t]
    \centering
    \begin{subfigure}[b]{\textwidth}
        \includegraphics[width=\textwidth]{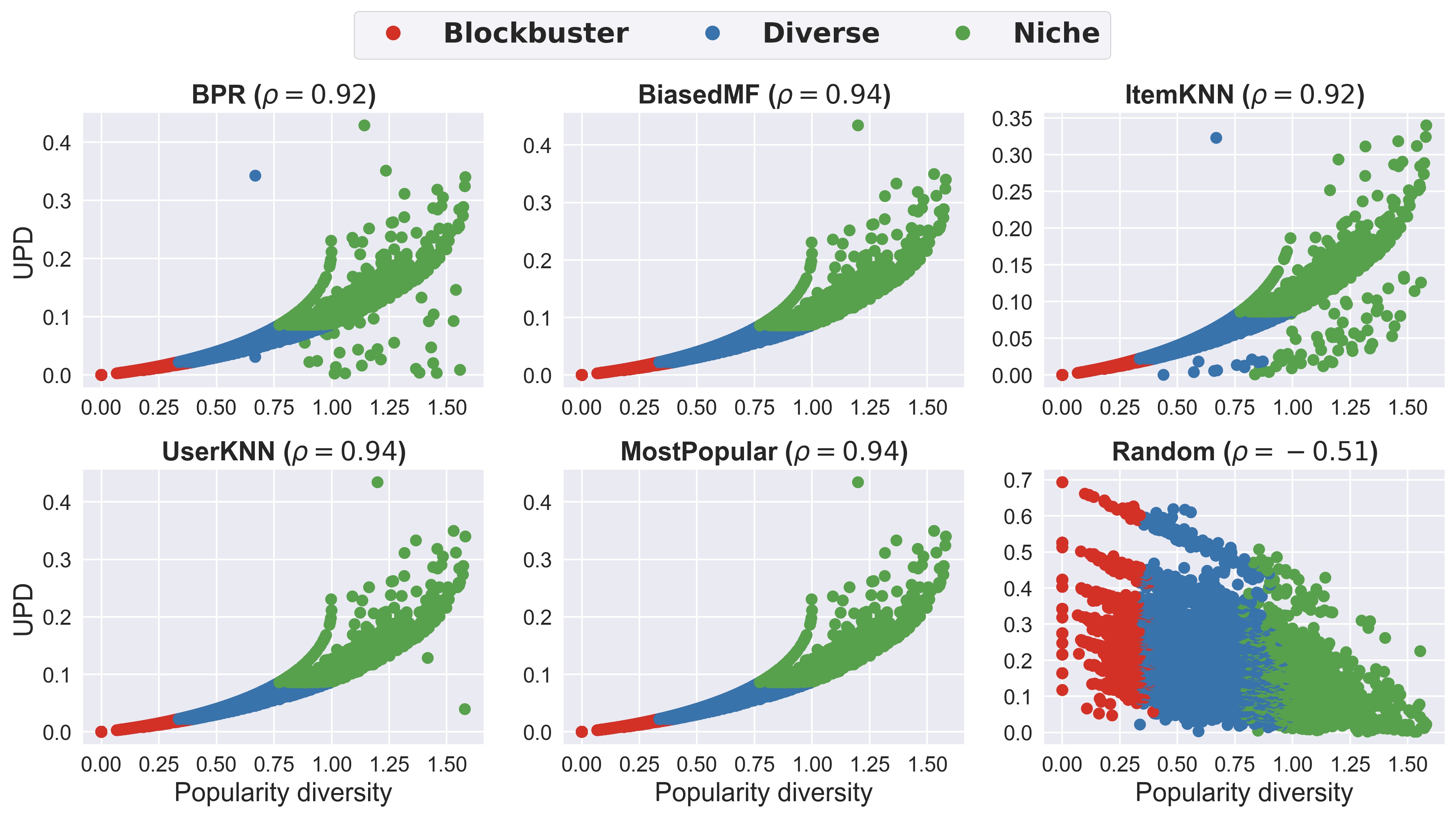}
    \end{subfigure}
    \vspace{-11pt}
\caption{The relationship between the popularity diversity of users' profile and $UPD$ of their recommendations on MovieLens dataset. $\rho$ is the correlation value of these two variables.}\label{fig_ml_pop_entropy_upd}
\end{figure*}

Figure \ref{fig_ml_pop_entropy_pl} shows the relationship between the popularity diversity and $PL$ of users' profile on each recommendation model. The horizontal axis is popularity diversity and vertical axis is $PL$. On each plot, the value $\rho$ shows the correlation between two variables. As shown, there is a positive correlation between popularity diversity and $PL$ of the users' profile. This correlation is higher on \algname{BPR}, \algname{BiasedMF}, \algname{UserKNN}, and \algname{Popular}. Analogously, Figure \ref{fig_ml_pop_entropy_upd} shows the relationship between the popularity diversity and $UPD$ of the users' profile on recommendation models. The same pattern can also be observed here, but the correlation is even higher (except for \algname{Random}).

While Figures \ref{fig_ml_pop_entropy_pl} and \ref{fig_ml_pop_entropy_upd} confirm that $PL$ and $UPD$ are correlated with popularity diversity of the users' profile, another pattern in these results is the popularity diversity across different user groups. It shows that blockbuster-focused group has the lowest popularity diversity, while niche-focused group has the highest popularity diversity. Possible explanation for this result is that blockbuster-focused users mainly interact with popular items, while niche-focused users interact with a combination of popular and unpopular items. 

Our last analysis reveals the relationship between the users' profile size and users' interest toward popular items. This analysis ensures that the correlation between our identified factors and the degree of unfairness is not due to some other hidden influential factors. The size of users' profile is an important factor that has been shown to affect the performance of recommendations (larger profile size leads to more accurate recommendation)~\cite{mansoury2020investigating}.

\begin{figure*}[t]
    \centering
    \begin{subfigure}[b]{0.8\textwidth}
        \includegraphics[width=\textwidth]{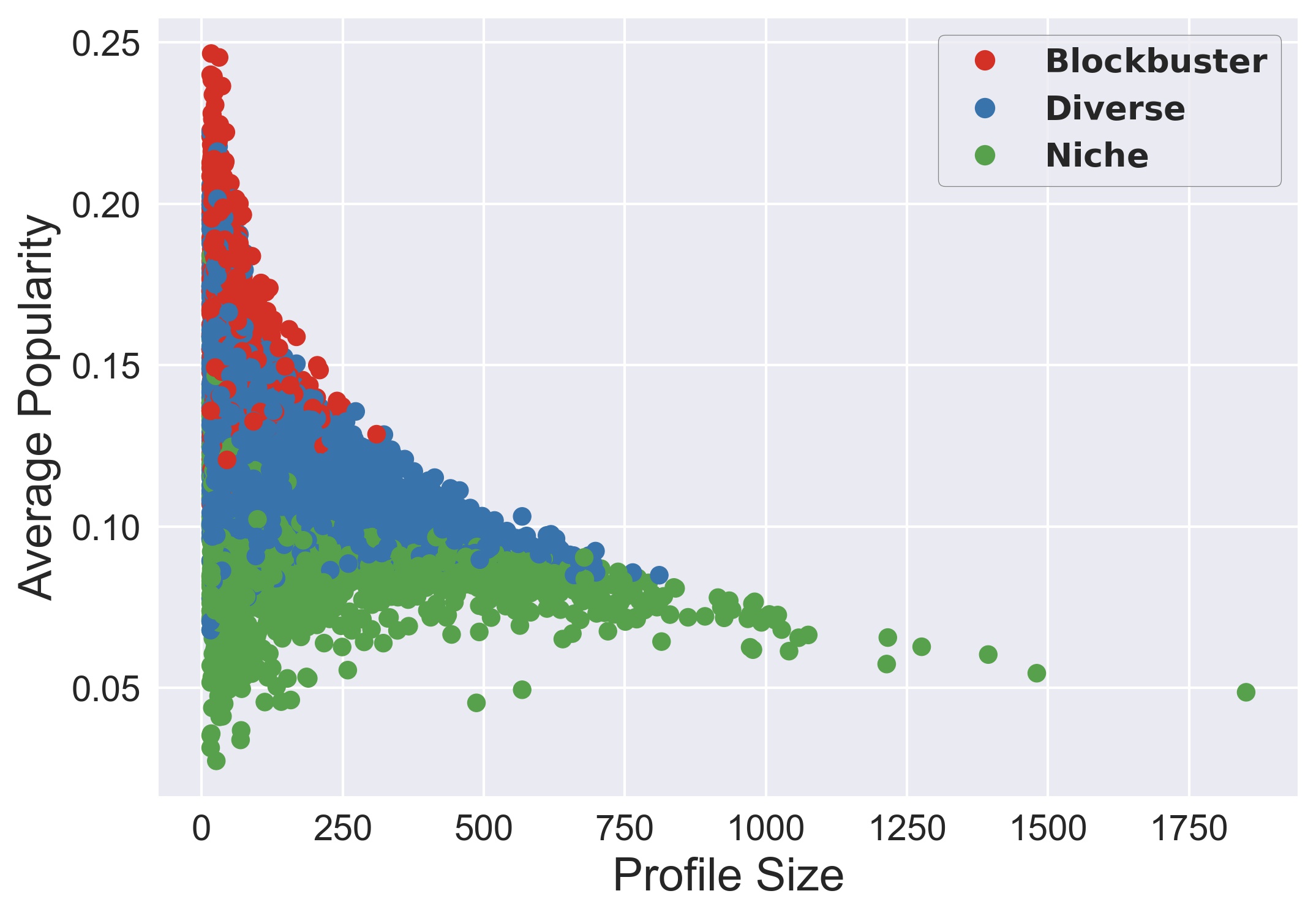}
    \end{subfigure}
    \vspace{-10pt}
\caption{The relationship between users' profile size and the average popularity of their profile.}\label{fig_ml_avg_pop_size}
\end{figure*}

Figure \ref{fig_ml_avg_pop_size} shows the relationship between average popularity and size of the users' profile for different groups of users. It is evident that $\textbf{Blockbuster}$ users have smaller profile (less interacted items), while $\textbf{Niche}$ users have larger profile size (more interacted items). With profile size being a factor affecting the quality of the recommendation for a user, the expectation is that the recommendations delivered to $\textbf{Niche}$ users better match their interest. However, our experimental results do not show this pattern. This confirms that our identified factors are not affected by users' profile size. 


\section{Conclusion}

In this paper, we studied the user-centered unfairness of popularity bias in recommender systems. In particular, we investigated two factors leading to this unfairness. One factor was profile inconsistency which refers to the degree of inconsistency of a user's profile in terms of her interest to item popularity and genre popularity (e.g., the inconsistency is high when a user is interested in popular items, but also interested in less popular genres). Another factor was popularity diversity of users' profile which refers to how diverse a user's profile is in terms of interacted item groups. Experiments on a movie recommendation dataset using six different recommendation algorithms showed that the introduced factors are highly correlated with the degree of unfairness in recommendation results. An interesting future direction is to investigate possible ways of utilizing the identified factors in this paper to improve the fairness of the recommendation models.

\bibliographystyle{splncs04}
\bibliography{ref}





\end{document}